\documentclass[final,5p,times,twocolumn]{elsarticle}

\usepackage{amssymb}

\usepackage{color}
\usepackage{amsmath}
\usepackage{multirow}
\usepackage{booktabs}
\usepackage{bm}
\usepackage{braket}
\usepackage{amsmath}
\usepackage{graphicx}
\biboptions{sort&compress}

\usepackage[pdfstartview=FitH, colorlinks,
 linkcolor={red!60!black!},
 anchorcolor={green!80!black!},
 urlcolor={blue!80!black!},
 citecolor={blue!50!black!}]{hyperref}
\usepackage[table]{xcolor}

\journal{Physics Letters B}

\begin{document}


\begin{frontmatter}

\title{Investigation of spectroscopic factors of deeply-bound nucleons in drip-line nuclei with the Gamow shell model}

\author[ad1,ad2]{M.R. Xie}
\author[ad1,ad2]{J.G. Li}
\author[ad1,ad2]{N. Michel\corref{correspondence}}
\author[ad1,ad2]{H.H. Li}
\author[ad1,ad2]{S.T. Wang}
\author[ad1,ad2]{H.J. Ong}
\author[ad1,ad2]{W. Zuo}

\address[ad1]{CAS Key Laboratory of High Precision Nuclear Spectroscopy, Institute of Modern Physics,
Chinese Academy of Sciences, Lanzhou 730000, China}
\address[ad2]{School of Nuclear Science and Technology, University of Chinese Academy of Sciences, Beijing 100049, China}

\cortext[correspondence]{Corresponding author. e-mail address: nicolas.michel@impcas.ac.cn (N. Michel)}

\begin{abstract}
Spectroscopic factors involving well bound nucleons in light nuclei are calculated with standard shell model, no-core shell model and Gamow shell model. Continuum coupling is included exactly 
in the Gamow shell model, due to the use of the Berggren basis, which contains bound, resonance and scattering states. Conversely, it is absent from standard and no-core shell models, where a basis of harmonic oscillator states is used. As the $A-1$ nuclei for which spectroscopic factors are calculated are either weakly bound or unbound, coupling to continuum is prominent, even though the $A$ nuclei are well bound. It is then showed that Gamow shell model can properly reproduce experimental data and is a predictive tool for detailed nuclear structure at drip-line, contrary to standard and no-core shell model.

\end{abstract}


\end{frontmatter}

 \section{Introduction}  
Light nuclei form unique laboratories to study nuclear structure in extreme conditions \cite{TANIHATA2013215,Motobayashi2014,BLANK2008403,RevModPhys.84.567}. As both proton and neutron drip-lines are experimentally accessible up to $A = 40$ \cite{wang2017ame2016,PhysRevLett.109.202503,Baumann20071022,PhysRevLett.122.052501,Thoenessen2004}, it is possible to study the nuclear interaction in weakly bound or resonance nuclei possessing an excess of protons or neutrons, respectively. 

Both proton-rich and neutron-rich systems in the light region of the nuclear chart exhibit similar behaviors, such as the presence of halos and particle-unbound nuclei of sizable widths at drip-line, i.e.~around 1 MeV or more, which can be easily reached \cite{ensdf}. It can occur that unbound proton-rich nuclei bearing $A < 10$ nucleons bear larger widths than their isobaric analog states. For example, $^5$He is unbound with a neutron-emission width of about 800 keV, whereas its proton-rich mirror $^5$Li has a proton-emission width of about 1.3 MeV \cite{ensdf}. This effect is even pronounced in the slightly heavier mirror systems $^7$He and $^7$B, bearing nucleon-emission widths of about 150 keV and 800 keV, respectively \cite{ensdf}. This is in contrast with medium and heavy nuclei at proton drip-line, which are confined by the Coulomb barrier and hence are very long-lived \cite{PhysRevC.56.1762,PhysRevLett.84.4549}. Many neutron halos, of one neutron or two-neutron character, have been identified in the nuclear chart: for example, $^{11}$Be, $^{15}$C, $^{19}$C and $^{23}$O for the former and $^{6}$He, $^{11}$Li, $^{14}$Be, $^{17}$B, $^{22}$C, $^{29}$F, and possibly $^{31}$F, for the latter \cite{Al-Khalili2004}. Proton halos also exist, even though they are in smaller number than neutron halos, e.g.~$^8$B, $^{13}$N, and the first excited state of $^{17}$F are one-proton halos, while $^{17}$Ne is a two-proton halo \cite{Al-Khalili2004}.

Besides the exotic structures of light nuclei at drip-lines, the simple adding of one proton in neutron-rich nuclei, or conversely, adding one neutron in proton-rich nuclei, strongly binds the newly obtained nuclear system. For example, while the two-neutron halo nucleus $^6$He is weakly bound by a little less than 1 MeV, $^7$Li has a neutron separation energy close to 7 MeV \cite{ensdf}. The same tendency is observed in their proton-rich mirrors, as the resonance $^6$Be has an energy of 1.4 MeV above the diproton-emission threshold, while the proton separation energy of $^7$Be is about 5.5 MeV \cite{ensdf}. Clearly, adding one nucleon to drip-line nuclei strongly modifies the many-body wave functions associated with these nuclei, in both the nuclear and asymptotic zones. Consequently, if one aims at representing many-body wave functions in these regions of the nuclear charts, models incorporating both inter-nucleon correlations and continuum coupling must be utilized. 

In order to assess the structure of many-body wave functions at drip-lines, overlap functions and spectroscopic factors (SFs) are very convenient quantities since they respectively represent the effect of adding a nucleon to a given $A-1$ nucleus in an $A$-nucleus in radial coordinate space and that integrated over the whole space. As a consequence, they are sensitive to the complex effects of inter-nucleon correlations on the asymptotic behavior of halo and resonance nuclei \cite{PhysRevC.75.031301,inpc2004,Michel_Springer,PhysRevC.104.L061301}. Nevertheless, theoretical calculations involving various models have shown discrepancies with experimental data (see Refs.\cite{PhysRevLett.93.042501,Gade_SFs} for extensive experimental studies of SFs on exotic nuclei). In fact, SFs are systematically overestimated in theoretical calculations \cite{PhysRevC.90.057602,PhysRevC.103.054610,PhysRevC.105.024613}.
However, the reason for this behavior is not well understood. Besides, the experimentally extracted SFs are always model-dependent, which results in additional uncertainty \cite{PhysRevLett.95.222501,KRAMER2001267,PhysRevC.65.034318,Hansen_Tostevin}. While short-range correlations are often thought to be responsible for SF overestimation \cite{POLLS1995371}, continuum coupling has shown to play a significant role in weakly bound and resonance nuclei \cite{PhysRevLett.107.032501,PhysRevC.75.031301,inpc2004,Michel_Springer}.

Here, we report on a theoretical investigation of overlap functions and SFs of deeply bound nucleons in drip-line nuclei using standard shell model (SM), no-core shell model (NCSM) and Gamow shell model (GSM). We applied the standard shell model (SM), defined in a single harmonic oscillator (HO) major shell, and hence very phenomenological for our purposes. We considered the NCSM, where all nucleons are active and interact via a realistic interaction, but where a localized basis of HO states is used. For GSM, one uses the Berggren basis, which contains bound, resonance and scattering states, and thus allows for an exact depiction of many-body nuclear wave functions asymptotes.

This paper is presented as follows. After shortly introducing the used models, namely SM, NCSM and GSM, we will depict the SFs and overlap functions calculated in considered nuclei. The differences in observables obtained with SM, NCSM and GSM will be emphasized, as well as the limitations of SM and NCSM to describe experimental data. Conclusion will be made afterwards.

\section{The Method} 

The many-body wave functions of considered nuclei are calculated using either SM, NCSM or GSM, hence with different Hamiltonian and model spaces. The NCSM Hamiltonian is defined from a realistic interaction, where all nucleons are active \cite{PhysRevLett.88.152502,PhysRevC.62.054311,PhysRevC.71.044312,PhysRevLett.99.092501,Navratil_2016}:
\begin{equation}
\hat{H}_{\rm NCSM} = \sum_{i=1}^A \frac{\mathbf{p}_i^2}{2m} - \frac{\mathbf{P}^2}{2mA} + \sum_{i<j}^{A}\hat{V}^{NN}_{ij},
\label{H_NCSM}
\end{equation}
where $m$ is the mass of a nucleon and $\hat{V}^{NN}$ is the used realistic two-body force. The many-body Schr{\"o}dinger equation $\hat{H}_{\rm NCSM} \ket{\Psi} = E \ket{\Psi}$ induced by Eq.~\ref{H_NCSM} is diagonalized using a basis of HO states. Center of mass excitations are removed using the standard Lawson method in $N \hbar \omega$ spaces.

NCSM is, in fact, the generalization of SM, the latter being typically defined in a single HO major shell \cite{COHEN19671,PhysRevC.74.034315,PhysRevC.69.034335}. The model space of SM is then that of a core plus valence nucleons.  In SM, one uses a phenomenological interaction fitted from experiment, where the Coulomb part is absent or renormalized in Hamiltonian parameters. One of the main advantages of SM is that center of mass excitations identically vanish in SM. Conversely, it is of very phenomenological use at drip-lines due to the very localized HO states forming its one-body basis.

GSM is based on the use of the one-body Berggren basis \cite{BERGGREN1968265}. The Berggren basis possesses bound, resonance and scattering states, which are generated by a finite-range potential, typically of Woods-Saxon (WS) type. It reads for a given partial wave of quantum numbers $\ell,j$:
\begin{equation}
\sum_n u_{n}^{(\ell j)}(r) u_{n}^{(\ell j)}(r') + \int_{L^+} u_{k}^{(\ell j)}(r)  u_{k}^{(\ell j)}(r') ~dk = \delta(r - r'), \label{Berggren}
\end{equation}
where $n$ enumerates the bound and resonance states of the considered partial wave, while $L^+$
is the complex contour of scattering states, which encompasses the resonance states present in the discrete sum. In order to use Eq.~\ref{Berggren} in numerical calculations, it is discretized with the Gauss-Legendre quadrature \cite{0954-3899-36-1-013101}:
\begin{equation}
\sum_n u_{n}^{(\ell j)}(r) u_{n}^{(\ell j)}(r') \simeq \delta(r - r'),  \label{Berggren_discr}
\end{equation}
where $n$ now depicts the bound, resonance and discretized scattering states of Eq.~\ref{Berggren}.
The many-body basis of GSM consists of the Slater determinants generated by the Berggren basis by occupying all possible one-body states of every proton or neutron partial wave \cite{0954-3899-36-1-013101,Michel_Springer,physics3040062}. Hence, the GSM Hamiltonian becomes a matrix to diagonalize, as in standard shell model. However, contrary to standard shell model, the GSM Hamiltonian matrix is complex symmetric \cite{0954-3899-36-1-013101,Michel_Springer}. Moreover, it possesses numerous many-body scattering states, so that the bound or resonance eigenstates are embedded among scattering eigenstates. Consequently, one had to develop numerical techniques to diagonalize the GSM Hamiltonian matrix. Thus, one developed the overlap method along with the Jacobi-Davidson method extended to complex-symmetric matrices to identify many-body resonance eigenstates \cite{0954-3899-36-1-013101,MICHEL2020106978,Michel_Springer}. Indeed, their energies are interior eigenvalues, and not extremal eigenvalues, as is the case for well bound states, whereby the use of the Lanczos method is optimal \cite{RevModPhys.77.427}.

Center of mass excitations in GSM cannot be corrected as in HO shell model, so that we introduced another framework for that matter. As the GSM Hamiltonian is defined in the core+valence particle picture, we can solve the many-body Schr{\"o}dinger equation within the so-called cluster orbital shell model (COSM) formalism \cite{PhysRevC.38.410} (see Refs.\cite{PhysRevC.84.051304,PhysRevC.96.054316,Michel_Springer} for several applications of COSM in GSM). For this, one defines the radial coordinates of valence nucleons with respect to the center of mass of the core. As a consequence, the coordinates of valence nucleons are translationally invariant, thereby suppressing all center of mass excitations by definition. The GSM Hamiltonian in COSM coordinates reads \cite{PhysRevC.84.051304,PhysRevC.96.054316,Michel_Springer}:
\begin{equation}
\!\!\!\hat{H}_{\rm GSM} \!=\!\! \sum_{i=1}^{A_{val}} \left( \frac{\mathbf{p}_i^2}{2 \mu_i} + \hat{U}_i^{(c)} \right) + \sum_{i<j}^{A_{val}} \left(  \hat{V}_{ij}^{(res)} + \frac{\mathbf{p}_i \cdot \mathbf{p}_j}{M_{c}} \right), \label{H_GSM}
\end{equation}
where $A_{val}$ is the number of valence nucleons, $\mu_i$ is the effective mass of the nucleon, $\hat{U}_i^{(c)}$ is the core potential acting on the $i$-th nucleon, $\hat{V}_{ij}^{(res)}$ is the residual inter-nucleon interaction, and the last term embodies the recoil effects induced by the finite mass of the core $M_{c}$ and is proper to the COSM formalism. Note that it would be very cumbersome to transform Eq.~\ref{H_NCSM} in COSM coordinates, that because of the different nature of core and valence coordinates. Hence, it is more convenient to define the GSM Hamiltonian in COSM coordinates, whose parameters are then directly fitted to experimental data \cite{PhysRevC.84.051304,PhysRevC.96.054316,Michel_Springer}. 

Overlap functions and SFs, denoted respectively by $O(r)$ and $C^2 S$, can then be defined similarly in SM, NCSM and GSM \cite{PhysRevC.75.031301,inpc2004,Michel_Springer}:
\begin{eqnarray}
O(r) &=& {\frac{1}{\sqrt{2J_A+1}}} \sum_n \braket{\Psi^{J_{A}}_{A} || a^+_{n \ell j} || \Psi^{J_{A-1}}_{A-1} } u_{n}^{(\ell j)}(r) \label{overlap_function}, \\
C^2 S &=& \int_0^{+\infty} O(r)^2~dr, \label{SF} 
\end{eqnarray}
where $\ell,j$ are the orbital and total angular momentum of the considered partial wave, respectively, $J_{A-1}$ and $J_A$ are the total angular momenta of the $A-1$ and $A$ nuclear wave functions $\ket{\Psi^{J_{A-1}}_{A-1}}$ and $\ket{\Psi^{J_A}_A}$, respectively, and where the $u_{n}^{(\ell j)}(r)$ are either the radial functions of the HO basis (SM and NCSM) or of the discretized one-body Berggren basis states of Eq.~\ref{Berggren_discr} (GSM).
Note that, in GSM, SFs are independent of the used single-particle basis. 

\section{Calculations and discussions} 
Let us firstly introduce the model space and interactions adopted in our SM, NCSM and GSM calculations, respectively. We used the standard SM framework with the Cohen-Kurath $p$-shell interaction  \cite{COHEN19671}. The realistic interaction used in NCSM is Daejeon16 \cite{SHIROKOV201687} and the model space is an $N \hbar \omega$ space, where the maximal value of $N$, denoted as $N_{max}$, runs from 0 to 10 and $\hbar \omega$ = 15 MeV. The Hamiltonian and model space in GSM are those of Ref.~\cite{PhysRevC.96.054316}. In GSM, the core is that of $^4$He, where its associated potential is of WS type and is fitted on the single-particle energies and phase shifts of nucleon + alpha direct reactions \cite{PhysRevC.96.054316}. The residual interaction is that introduced by Furutani, Horiuchi and Tamagaki (FHT) \cite{FHT1,FHT2}, to which the Coulomb interaction is added in the proton valence part. The valence space consists of the $spdf$ partial waves of proton and neutron types, where the $spd_{5/2}$ partial waves are represented by the Berggren basis, with the contour of scattering states discretized with 30 points for each partial wave, while the $d_{3/2}f$ partial waves are each represented by six HO states ($n=0,\dots,5$). Truncations are imposed in GSM, where one demands at most three occupied states in the non-resonant continuum spanned by the Berggren basis.

\begin{figure}[!htb]
    \centering
    \includegraphics[width=1.\columnwidth]{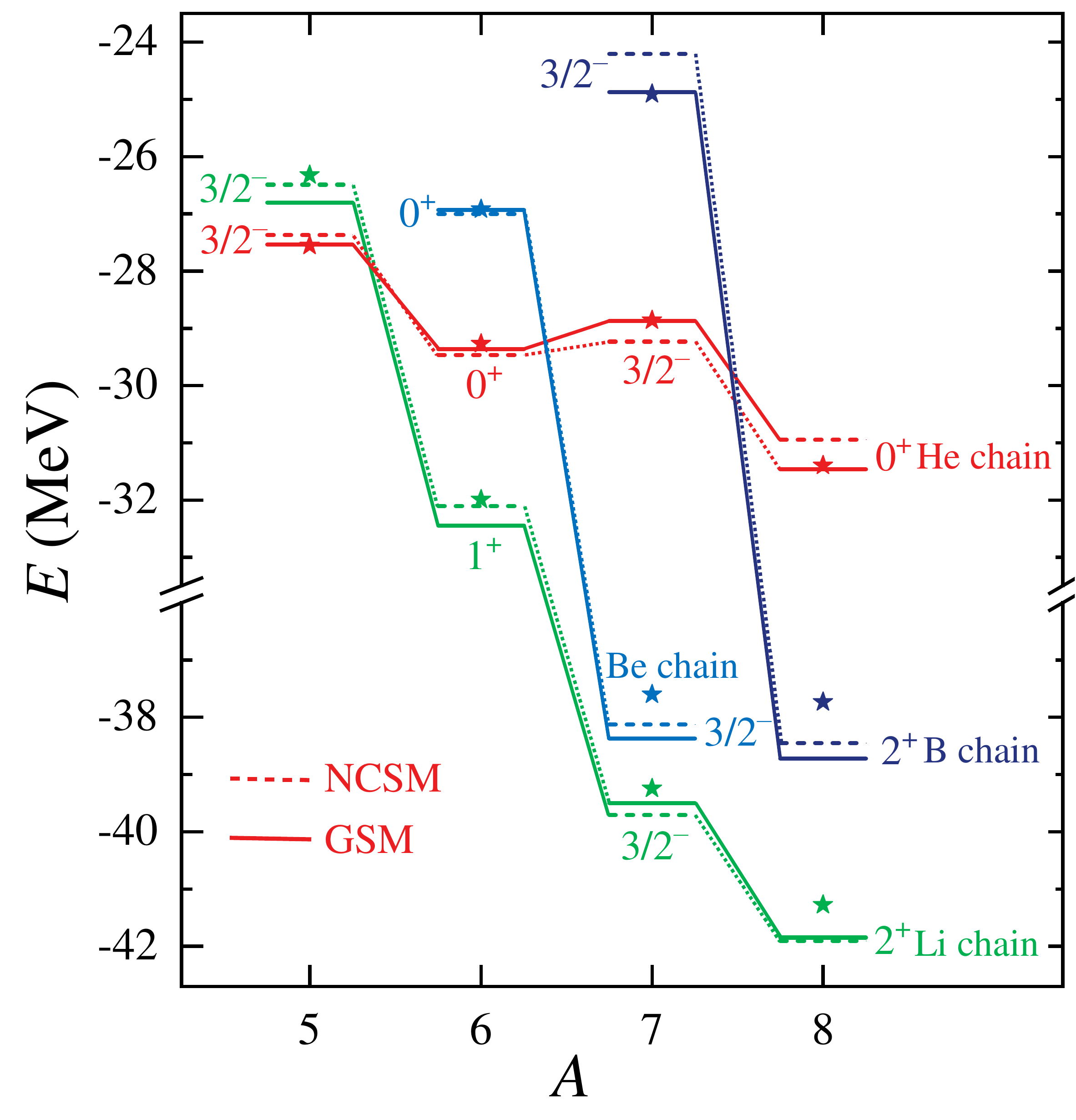}
    \caption{Low-lying spectra of the isotopes of He, Li, Be and B bearing $A=7,8$ nucleons calculated with NCSM (linked with dotted lines) and GSM (linked with solid lines). Experimental data are taken from Ref.\cite{ensdf} and are indicated by stars.}
    \label{fig:energy}
\end{figure}

We firstly compare the spectrum of the light nuclei of interest obtained in NCSM and GSM with experimental data in Fig.~\ref{fig:energy}. One can see that both NCSM and GSM satisfactorily reproduce experimental data, as the typical error is about 500 keV to 1 MeV. Typically, GSM energies are closer to experimental data than NCSM, because the FHT interaction has been directly fitted to the isotopes of He, Li and Be chains. However, as we can notice for the ground states of $^7$Be, $^8$B and $^8$Li, NCSM and GSM ground state energies are almost the same. Consequently, the use of realistic interactions in a no-core approach or effective interactions fitted in the core+valence particles picture leads to more or less the same quality of results when considering energy spectra. However, this does not provide any insight about many-body wave functions, especially about their asymptotic behavior.

A very interesting observable for that matter is the overlap function (see Eq.~\ref{overlap_function}). As it is a function of the radial coordinate, it provides information about nuclear structure inside the nucleus and also in the asymptotic zone \cite{PhysRevC.85.064320,LI2022137225}.
Indeed, an overlap function will be extended in space if the $A-1$ or $A$ nucleus present in Eq.~\ref{overlap_function} is weakly bound or resonant. This is obvious if the $A$ nucleus lies close to particle-emission threshold and the $A-1$ nucleus is well bound. In this case, the overlap function can be seen, at least qualitatively, as a halo or resonance single-particle wave function associated to the added nucleon. On the contrary, if the $A-1$ nucleus is weakly bound or resonant and the $A$ nucleus bears a relatively large binding energy, the extended overlap function is affected by the cancellation of the extended density of the $A-1$ nucleus by the added nucleon. Indeed, the surplus of binding energy provided by the added nucleon localizes the density of the $A$ nucleus. This effect must then necessarily be seen in the overlap function, as otherwise this would imply that the $A-1$ and $A$ nuclear wave functions have similar asymptotes.

\begin{figure}[!htb]
    \centering
    \includegraphics[width=1.\columnwidth]{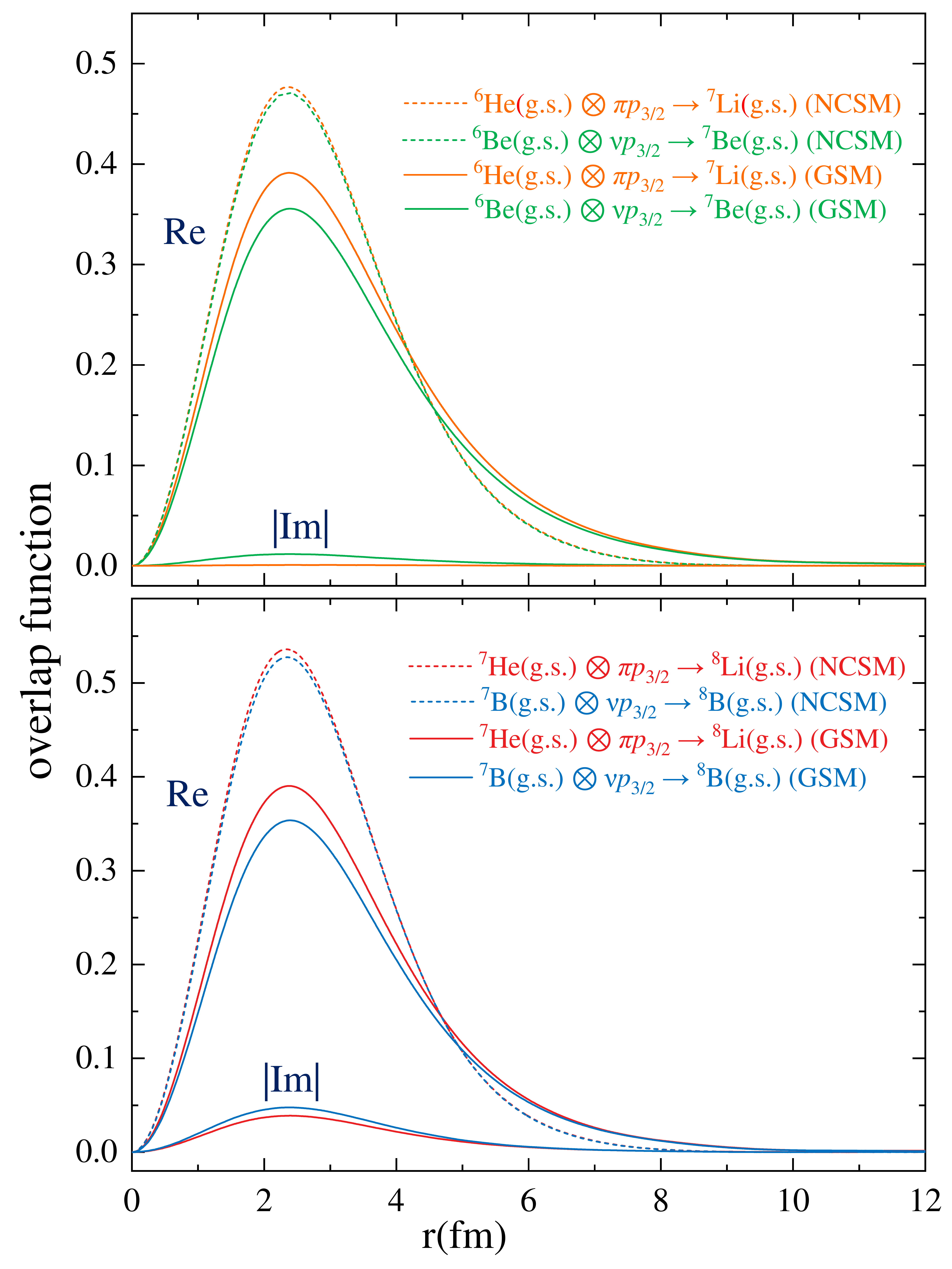}
    \caption{Proton and neutron overlap functions of the systems $\ket{A=6} \otimes p_{3/2} \rightarrow \ket{A=7}$ (upper) and $\ket{A=7} \otimes p_{3/2} \rightarrow \ket{A=8}$ (lower) calculated with NCSM and GSM models, where isotopes belong to the He, Li, Be and B chains. 
      Both the real (Re) and absolute value of imaginary parts ($|$Im$|$) of GSM overlap functions are shown. Only nuclear ground states (g.s.) are considered.
      Imaginary parts in absolute value represent the statistical error made on real parts due to the finite life-time of the parent nucleus \cite{BERGGREN1968265,BERGGREN19961,Michel_Springer} (color online).}
    \label{fig:overlap-functions}
\end{figure}

Our calculated overlap functions involving $p_{3/2}$ partial waves in He, Li, Be and B isotopes bearing $A=6-8$ nucleons are depicted in Fig.~\ref{fig:overlap-functions}. The $p_{3/2}$ partial wave is indeed dominant in these nuclei as the $0p_{3/2}$ proton or neutron states are the last occupied single-particle states at independent-particle level. One can see that the overlap functions obtained in GSM are all extended in space. This is the case because the $A-1$ nucleus is always weakly bound ($^6$He) or a many-body resonance ($^6$Be, $^7$He, $^7$B), while the $A$ nucleus is either well bound ($^{7,8}$Li, $^7$Be) or weakly bound ($^8$B). Indeed, the effect of the proton halo of $^8$B in the overlap function associated to the $^7$B + n $\rightarrow$ $^8$B reaction is relatively small, because the rather broad resonance of $^7$B comparatively generates a larger overlap function strength in the asymptotic region.

In this context, one can note that non-zero imaginary parts in calculated overlap functions arise because nuclear parent states are unbound. They provide the statistical error on the real parts of overlap functions generated by the finite life-time of parent nuclei \cite{BERGGREN1968265,BERGGREN19961,Michel_Springer}. One can see that they are the largest for the $^7$B + n $\rightarrow$ $^8$B and $^7$He + p $\rightarrow$ $^8$Li reactions and that it is smaller, but still visible, for the $^6$Be + n $\rightarrow$ $^7$Be reaction. It could be expected that the smallest imaginary parts in absolute values occur for the $^6$Be + n $\rightarrow$ $^7$Be reaction, as $^6$Be is the narrowest considered resonance, with a width of about 90 keV. However, the absolute values of the imaginary parts of the GSM SF of the $^7$B + n $\rightarrow$ $^8$B and $^7$He + p $\rightarrow$ $^8$Li reactions are close, even though the respective widths of $^7$He and $^7$B are about 200 keV and 770 keV in our calculations, to be compared to their respective experimental value of about 150 and 800 keV. This shows that the imaginary part of overlap function is not proportional to the width of its associated parent nucleus, even though it is expected to increase in absolute value when the parent nucleus becomes more unbound, as it occurs in our calculations. Indeed, nuclear structure also plays a role in the quantitative values of overlap functions. As $^7$He and $^7$B are mirror nuclei, it is the Coulomb interaction, present only in $^7$B, which is responsible for their different nuclear structure. Indeed, the Coulomb interaction not only modifies the energies of $^7$He and $^7$B with respect to particle-emission threshold and their widths, but also their wave functions via isospin-symmetry breaking.
 
Conversely, all NCSM overlap functions are localized. They decay faster than GSM overlap functions in the asymptotic region and are also larger in the nuclear zone. This occurs because $A$ nuclear wave functions are normalized: indeed, if particle density decreases in the asymptotic region in a partial wave where nucleons are very likely to be found, it must increase in the nuclear zone for particle number to be conserved. Consequently, GSM overlap functions reflect properly the asymptotes of $A$ and $A-1$ nuclei, contrary to those arising from NCSM. As SFs are equal to the norm of overlap functions (see Eq.~\ref{SF}), one can then expect the SFs obtained in GSM to better reproduce experimental data than those of NCSM.

SFs, strictly speaking, are not observables in nuclear physics \cite{SF_non_observable,Liu_2020,Furnstahl_2010}. Contrary to phase shifts, energies and asymptotic normalizations, they do not arise from the S-matrix and cannot be part of a standard reaction theory \cite{SF_non_observable}. SFs, however, can be observables in situations where typical momentum scales are small, such as in cold atoms \cite{Furnstahl_2010}. Added to that, it is not clear whether the non-observable character of SFs in nuclear physics arises from the long-range or short-range parts of the interactions, respectively associated to low and high momentum components in nuclear wave functions \cite{Furnstahl_2010}. In fact, SFs are relevant to the renormalization scheme used for Hamiltonian definition and explicitly depend on the cut-off energy defining the nuclear interaction used, for example. Consequently, the experimental extraction of SFs is model-dependent and hence subject to debate \cite{PhysRevLett.95.222501,KRAMER2001267,PhysRevC.65.034318,Hansen_Tostevin}. Nevertheless,  SFs can provide useful information about nuclear structure and inter-nucleon correlations. Indeed, a large SF in a given partial wave implies that one nucleon is largely decoupled from the other $A-1$ remaining nucleons, whereas a small value for SF indicates that important inter-nucleon correlations occur in the considered partial wave. 

\begin{figure}[!htb]
    \centering
    \includegraphics[width=1.\columnwidth]{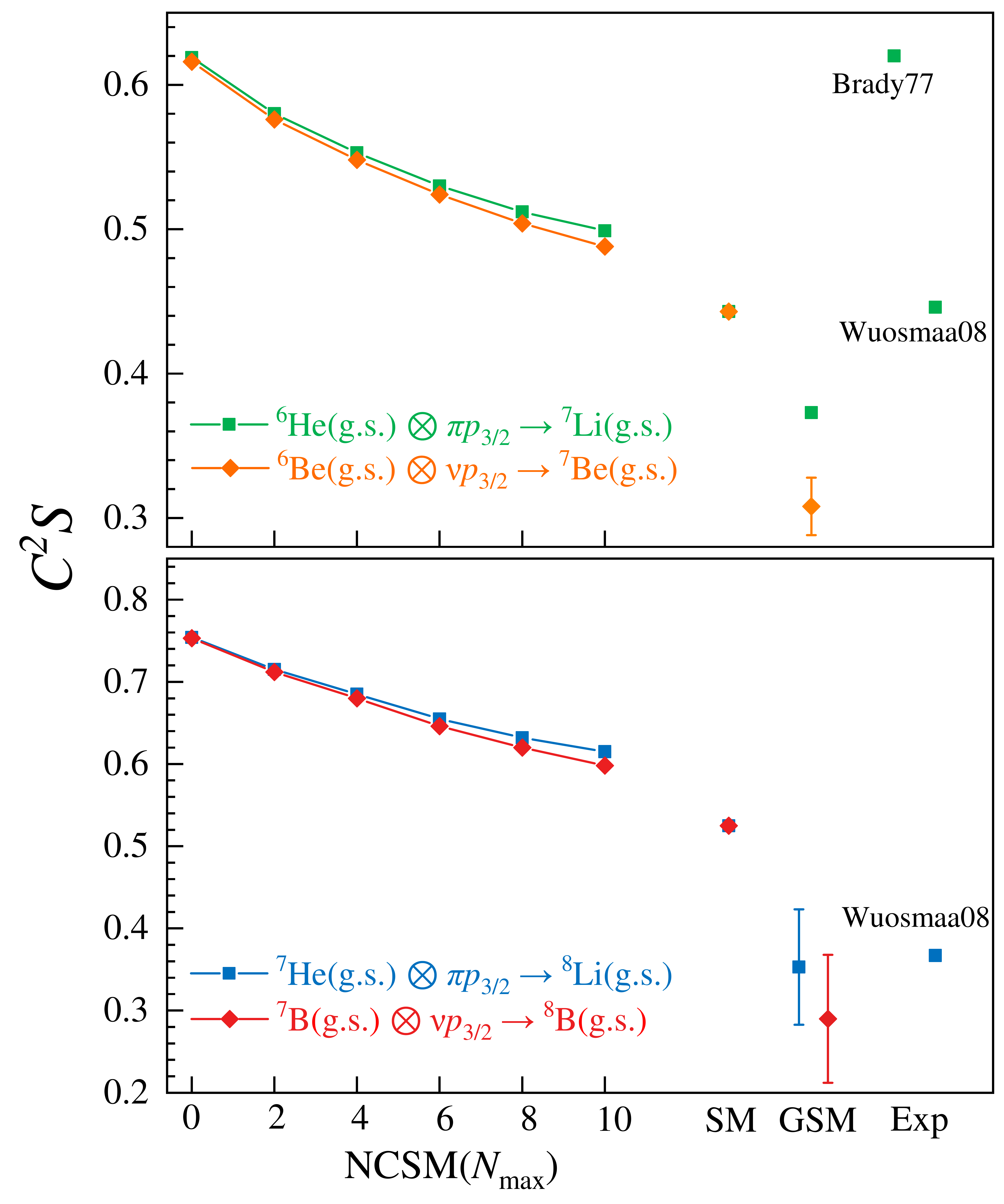}
    \caption{SFs associated to the overlap functions shown in Fig.~\ref{fig:overlap-functions}. The models used for their evaluations are SM, NCSM and GSM. Proton overlap functions results are illustrated by squares and neutron overlap functions by lozenges. The absolute values of the imaginary parts of GSM SFs are represented by error bars as they correspond to the statistical errors made on GSM SFs real parts due to the finite life-time of the parent nucleus \cite{BERGGREN1968265,BERGGREN19961,Michel_Springer}. The values of SFs obtained in NCSM are explicitly depicted as a function of the $N_{max}$ of the $N \hbar \omega$ space. Experimental data (Exp) are taken from Refs.\cite{PhysRevC.16.31,PhysRevC.78.041302}.  }
    \label{fig:SF}
\end{figure}

Calculations of the SFs associated to the overlap functions of Fig.~\ref{fig:overlap-functions} performed in NCSM and GSM, as well as SM in $p$-space, are illustrated in Fig.~\ref{fig:SF} and compared to experimental data.
One can see in Fig.~\ref{fig:SF} that NCSM is not reliable for the calculations of SFs in nuclei close to particle-emission threshold. Indeed, the calculated NCSM SF values are not converged even when using large model space dimensions. They continue to significantly decrease in the $10 \hbar \omega$ space, which is the largest space available in our calculations, and one can see from the SF values obtained in $0,\dots,8 \hbar \omega$ spaces that convergence is far from being reached. In fact, it is not even guaranteed that convergence will take place when unbound nuclei enter SFs. Indeed, $^6$Be, $^7$He and $^7$B are all resonances, so that they cannot be expanded with a basis of HO states \cite{PhysRevC.103.034305}. While $^6$He is bound and can be represented in an HO basis in principle, a similar difficulty applies to its many-body wave function in practice. A very large number of HO states would be required to depict its halo in the asymptotic zone. Clearly, the model space dimensions of the $N \hbar \omega$ spaces used in NCSM able to depict halo structure bear extremely large values, which are so far out of reach computationally. Thus, it is very likely that NCSM cannot encompass the halo and unbound nuclear structure necessary to accurately provide SFs associated to drip-line nuclei.

Conversely, as the $p_{3/2}$ partial wave is fully considered in GSM, we can directly obtain precise SFs in GSM. GSM SFs compare well with available experimental data. The absolute values of imaginary parts of shown GSM SFs provide information on the statistical uncertainty of GSM SFs real parts generated by the finite life-time of parent nuclei \cite{BERGGREN1968265,BERGGREN19961,Michel_Springer}. They are represented by error bars on Fig.~\ref{fig:SF}. It can be seen that the imaginary parts of GSM SFs have the same properties as those of their associated overlap function (see discussion above about imaginary parts of overlap functions). Surprisingly, the SM SF associated to the $^6$He + p $\rightarrow$ $^7$Li provides the closest value to experimental data. But this might be accidental. Indeed, the SM SF arising from the $^7$He + p $\rightarrow$ $^8$Li reaction is rather far from experimental data, whereas the GSM SF is very close to the experimental value. In addition, the SM SFs arising in mirror systems are identical because the Coulomb interaction is absent in SM. As the difference between the SFs arising from mirror systems is significant in GSM, it is clear that the mirror SFs cannot be properly described in SM. Model spaces are larger in NCSM, with the Coulomb interaction also present in the NCSM Hamiltonian. However, the quality of results in NCSM is comparable to that of SM. 
Unlike in SM, the NCSM SFs of mirror systems are slightly different, but markedly larger than those of GSM. Hence, as many-body wave function asymptotes are properly described in GSM, one can expect that the SFs values obtained with GSM are the most precise. Moreover, GSM results show that isospin-symmetry breaking also occurs in SFs and overlap functions of the investigated mirror partners. For that matter, one can see in Fig.~\ref{fig:SF} that the reduction of neutron SFs in GSM with respect to those of SM and NCSM is stronger than that of GSM proton SFs. This additional quenching originates from the presence of the infinite-range Coulomb Hamiltonian in the unbound parent nucleus when neutron SFs are considered. Indeed, the Coulomb Hamiltonian generates an additional continuum coupling in the asymptotic region therein, which does not exist in the neutron-unbound GSM parent nuclei involved in GSM proton SFs. Thus, the stronger reduction of neutron SFs in GSM compared to GSM proton SFs is a feature of isospin-symmetry breaking. 

Conversely, SFs and overlap functions are very close in NCSM calculations (see Figs.(\ref{fig:overlap-functions},\ref{fig:SF})). Hence, our results clearly demonstrate that continuum coupling is indispensable for the proper description of isospin-symmetry breaking effects in the SFs of dripline nuclei. This conclusion goes along the line of the results obtained in Ref.\cite{PhysRevLett.107.032501}, where the continuum effects of SFs in the removal of deeply-bound nucleons are discussed within coupled-cluster calculations. Indeed, as in Ref.\cite{PhysRevLett.107.032501}, our SFs are smaller than those obtained with methods relying on HO bases (see Fig.~\ref{fig:SF}). This shows that SF quenching is independent of the realistic or phenomenological character of the used Hamiltonian and that continuum coupling is prominent for that matter. Indeed, realistic interactions have been used in Ref.\cite{PhysRevLett.107.032501}, whereas our nuclear Hamiltonian consists of a core WS potential and FHT residual interaction, fitted on experimental data.

A precise calculation of SFs is necessary to evaluate particle-removal cross sections. Indeed, these cross sections are typically modeled with the following equation \cite{PhysRevC.90.057602}:
\begin{equation}
\sigma^{(\ell j)} = C^2 S_{\ell j} ~ \sigma^{(\ell j)}_{s.p.}  \ ,
\label{cross_section_from_SF}
\end{equation}
where $\sigma^{(\ell j)}$ is the cross section of the removal of a nucleon in the $\ell j$ partial wave, $C^2 S_{\ell j}$ is the associated SF of Eq.~\ref{SF} and $\sigma^{(\ell j)}_{s.p.}$ is the single-particle (s.p.) cross section, containing the kinetic dependence of the transfer reaction and typically provided by an optical potential. As could be clearly seen from Fig.~\ref{fig:SF} (see also Refs.~\cite{PhysRevC.90.057602,PhysRevC.103.054610,PhysRevC.105.024613}), the major deficiency of Eq.~\ref{SF} in practical calculations is that $C^2 S_{\ell j}$ values are either calculated in SM using a single major HO shell or lack continuum coupling in NCSM because they are evaluated from a one-body HO basis. Consequently, the cross sections evaluated from Eq.~\ref{cross_section_from_SF} will be closer to experimental data if SFs are calculated in a full $\ell j$ partial wave space including continuum coupling, as is performed in GSM.

\section {Summary} 
The nuclear structure of drip-line nuclei is not well understood. SFs, which reflect inter-nucleon correlations, are typically overestimated in current nuclear models. Besides the systematic errors encountered in experiments, the theoretical inadequacies are often thought to originate from short-range correlations. However, they might also arise form continuum coupling, which is important in drip-line nuclei and is often overlooked in realistic nuclear calculations.

Consequently, we performed calculations of overlap functions and SFs in drip-line nuclei bearing $A=6-8$ nucleons using several nuclear models. For this, we applied SM, NCSM and GSM, as they typically reproduce experimental spectra. Among the considered models, continuum coupling is explicitly included in GSM via the use of the Berggren basis, so that the effects of the nuclear asymptotes of halo and resonance states can directly be assessed.

It has then been noticed from our calculations that SM and NCSM, whose many-body basis is generated from HO states, are inadequate for a proper calculation of overlap functions and SFs involving drip-line nuclei. Indeed, SM cannot differentiate between mirror systems as the Coulomb interaction is absent from its Hamiltonian, while SFs do not converge with increasing $N_{max}$ in NCSM. This occurs because halo states exhibit slow convergence when expanded with a HO basis, and, from a strict theoretical point of view, because resonances cannot be represented with a HO basis. On the contrary, converged results are obtained with GSM, as the Berggren basis allows to reproduce nuclear asymptotes of both weakly bound and unbound states. We could then show that the SFs obtained with GSM are smaller than those arising from SM and NCSM and agree with experimental data. 

Our results show that the inclusion of continuum coupling is necessary to explain the SF discrepancy between theory and experiment. As GSM makes use of an effective interaction fitted on experimental energies in a core+valence nucleon picture, our results suggest that continuum coupling is more important than the details of the used nucleon-nucleon interaction to properly depict experimental SFs. Calculations including both realistic interactions and continuum coupling would provide more information for that matter. Although a no-core approach using the Berggren basis has been developed and successfully applied to very light isotopes \cite{PhysRevC.88.044318,PhysRevC.100.054313,PhysRevC.104.024319}, calculations become unfeasible in practice when approaching $A \sim 6,7$. Hence, the use of the newly developed ab-initio GSM with a core
\cite{HU2020135206,ZHANG2022136958} would be helpful, as it combines realistic framework and the convenience of the core+valence nucleon approach.

\textit{Acknowledgments} -- 
 We thank Chunwang  Ma for his suggestions and useful comments.
 This work has been supported by the National Natural Science Foundation of China under Grant Nos. 11975282, 12205340, and 12175281; the Strategic Priority Research Program of Chinese Academy of Sciences under Grant No. XDB34000000; the Key Research Program of the Chinese Academy of Sciences under Grant No. XDPB15; the State Key Laboratory of Nuclear Physics and Technology, Peking University under Grant No. NPT2020KFY13. This research was made possible by using the computing resources of Gansu Advanced Computing Center.

\section*{References}

\bibliographystyle{elsarticle-num_noURL}
\bibliography{Ref}





\end{document}